\documentclass[12pt]{article}

\usepackage{graphicx}
\usepackage{dsfont}
\usepackage{amssymb}
\usepackage{mathrsfs}

\begin{document}
\title{Notes on Conformal Invisibility Devices}
\author{Ulf Leonhardt\\
School of Physics and Astronomy, University of St Andrews,\\
North Haugh, St Andrews KY16 9SS, Scotland
}
\date{\today}
\maketitle
\begin{abstract}
As a consequence of the wave nature of light,
invisibility devices based on isotropic media cannot be perfect.
The principal distortions of invisibility
are due to reflections and time delays.
Reflections can be made exponentially small for devices
that are large in comparison with the wavelength 
of light.
Time delays are unavoidable and 
will result in wave-front dislocations.
This paper considers invisibility devices based
on optical conformal mapping.
The paper shows that the time delays
do not depend on the directions and impact parameters
of incident light rays,
although the refractive-index profile
of any conformal invisibility device is
necessarily asymmetric.
The distortions of images are thus uniform,
which reduces the risk of detection.
The paper also shows how the ideas of 
invisibility devices are
connected to the transmutation of force,
the stereographic projection
and Escheresque tilings of the plane.\\

\end{abstract}

\newpage

\section{Introduction}

\begin{figure}[h]
\begin{center}
\includegraphics[width=17.0pc]{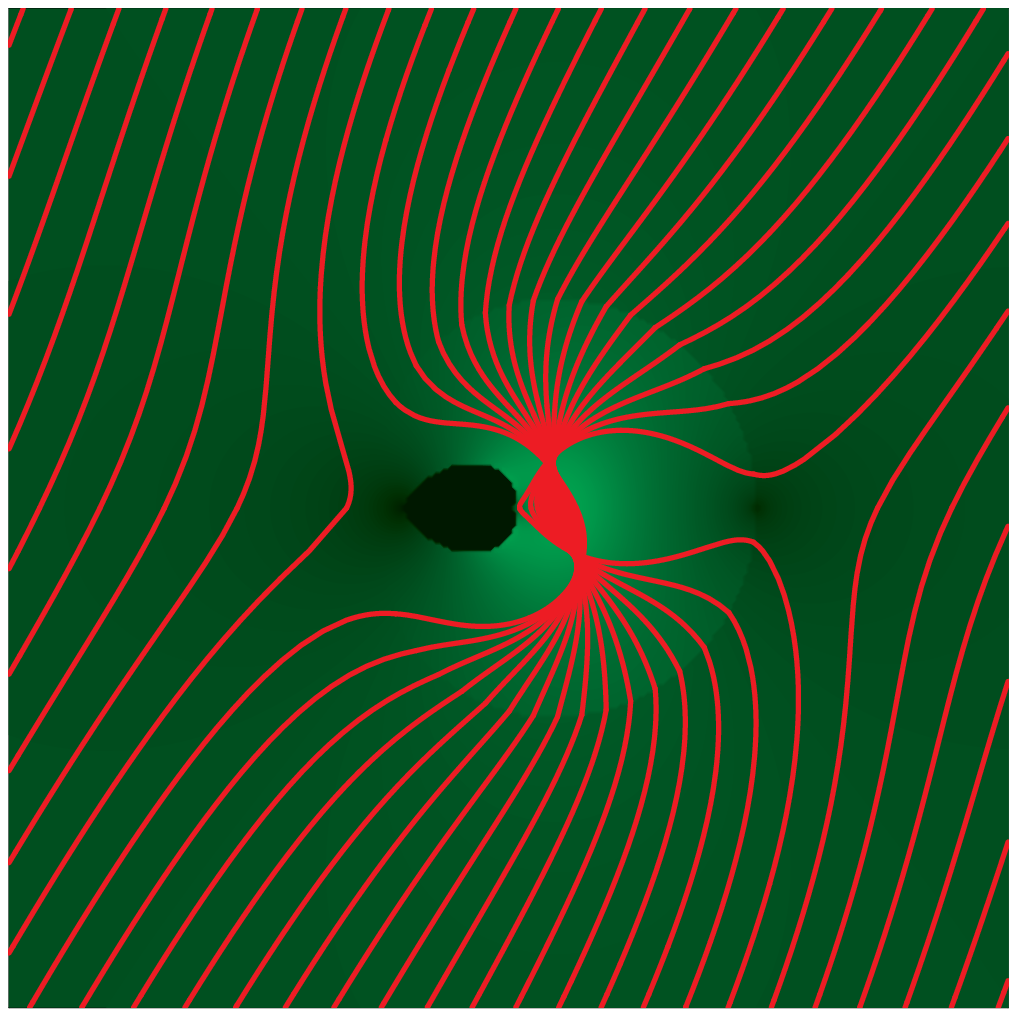}
\includegraphics[width=17.0pc]{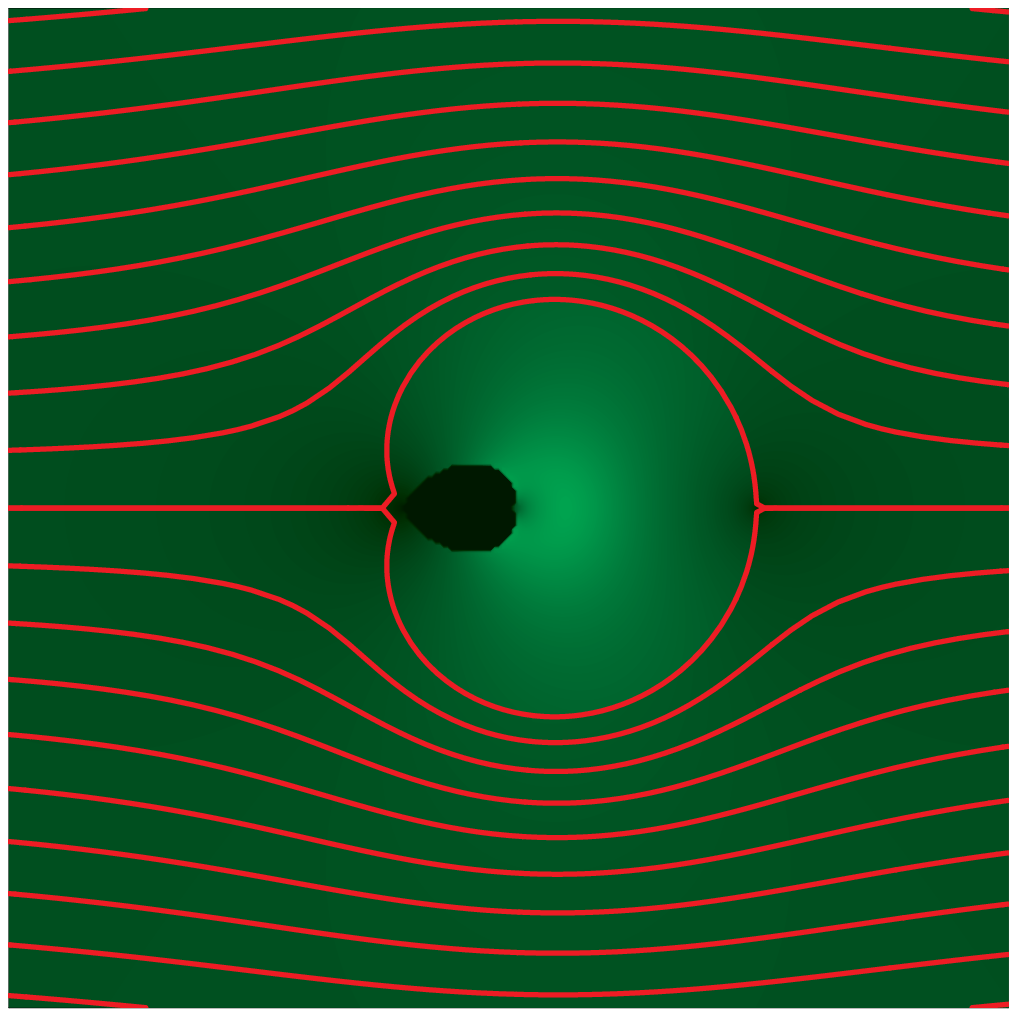}
%\vspace*{-15mm}
\caption{{\small
Light propagation in a conformal invisibility device.
The light rays are shown in red. 
The brightness of the green background 
indicates the refractive-index profile. 
The device is based on the conformal map 
(\ref{eq:wmap}) and 
Maxwell's fish eye 
(\ref{eq:maxwelleye})
with $r_3=4r_0$ and $n_0=2$
(further details given in the Appendix).
The device consists of an exterior
and an interior layer with a clear boundary
illustrated by the tiling in Fig.\ \ref{fig:surface}.
The invisible region is shown in black.
Anything could be placed there.
The left figure illustrates how light is refracted at
the boundary between the two layers
and guided around the invisible region
where it leaves the device as if nothing were there.
In the right figure, 
light simply flows around the interior layer.
}
\label{fig:flow}
%\vspace*{-15mm}
}
\end{center}
\end{figure}

The bending of light in dielectric media \cite{BornWolf}
is the cause of many optical illusions.
For example, in a mirage in the desert \cite{Feynman},
light rays from the sky
are bent above the hot sand where the air is thin
and the refractive index is low.
In this way the rays minimize their optical paths
according to Fermat's Principle \cite{BornWolf}.
They are creating images of the sky that deceive the observer
as illusions of water \cite{Feynman}.
Imagine a different situation 
\cite{Gbur,Invisibility,LeoConform}
where a medium guides light around a hole in it
such that the rays leave the medium as if nothing were there,
see Fig.\ \ref{fig:flow}.
Any object placed inside would be hidden from sight.
The medium would create 
the ultimate optical illusion: invisibility \cite{Gbur}.
Recently, 
ideas for designing such invisibility devices
have been discussed \cite{Invisibility,LeoConform}.
Ideas for minuscule invisible bodies
(smaller than the wavelength of light)
are older \cite{Kerker},
but first schemes for implementations and 
interesting twists and methods 
have been developed recently
\cite{Alu,Milton}.
Cloaking devices require unusually strong
refractive-index profiles,
but it is conceivable that they can be built with
dielectric metamaterials \cite{Alu,Milton,Smith}.
Such devices 
would operate in the microwave region of the 
electromagnetic spectrum \cite{Smith}
and perhaps also in some frequency windows in the 
visible range \cite{Grigorenko}.

%\vspace*{-10mm}
%%%
%%%

\newpage

Strictly speaking, ideal invisibility devices 
based on isotropic media are 
impossible due to the wave nature of light 
\cite{Nachman,WolfHabashy}.
Highly anisotropic media, however,
may lead, in principle,
to the construction of perfect invisibility devices \cite{Invisibility}.
Expressed in mathematical terms, 
the inverse scattering problem for linear waves 
in isotropic media has unique solutions \cite{Nachman}.
Therefore, the asymptotic behavior 
of propagation through empty space, 
or a uniform medium, is only consistent with
the actual propagation through a uniform medium.
In theory, nothing can be hidden.
In practice, a dielectric invisibility device
would perhaps create a slight haze, instead of a perfect image.
The principal distortions of invisibility
are due to reflections and time delays.
Reflections can be made exponentially small for devices
that are large in comparison with the wavelength 
of light \cite{LeoConform}.
Time delays are unavoidable.
Unless the phase delays are tuned to be multiples of $2\pi$
for some certain frequencies,
they will result in wave-front dislocations 
at boundaries that lead to image distortions
due to diffraction \cite{BornWolf}.
Wave-front dislocations also pose the risk of detection 
by sensitive wave-front sensors \cite{Tyson}.
In this paper we calculate the time delay 
caused by the scheme \cite{LeoConform}
based on optical conformal mapping,
assuming two-dimensional light propagation 
from infinity.
We find that the delay is uniform for all directions
and impact parameters,
although the refractive-index profiles of 
invisibility devices are necessarily asymmetric \cite{Awatif}.
Therefore,
the distortions of images composed of 
various spatial Fourier components are uniform,
which reduces the risk of detection.

\section{Theory}

Our theory is based on geometrical optics \cite{BornWolf}
and in particular on 
Fermat's Principle \cite{BornWolf} and on
Hamilton's analogy \cite{BornWolf}
between the propagation of light in media and the motion of 
particles in classical mechanics \cite{LL1}.
Suppose that the refractive index profile $n(\mathbf{r})$ 
does not vary
much over scales comparable with the wavelength of light.
In this regime of geometrical optics
both polarization components of light $\psi$
for frequencies $\omega$
obey the Helmholtz equation \cite{BornWolf}
%%%%%%
\begin{equation}
\left(\nabla^2 + n^2 \frac{\omega^2}{c^2}\right) \psi = 0
\label{eq:helmholtz}
\end{equation}
%%%%%%
where $c$ denotes the speed of light in vacuum.
The Helmholtz equation (\ref{eq:helmholtz}) 
is equivalent to the stationary Schr\"odinger equation
with potential $U$ and energy $E$ such that \cite{BornWolf}
%%%%%%
\begin{equation}
U-E = -\frac{n^2}{2} \,.
\label{eq:potential}
\end{equation}
%%%%%%
Therefore we expect that Hamilton's equations for light rays
are equivalent to Newton's equations 
of mechanical particles\footnote{Light rays in moving media 
behave like particles in magnetic fields for low
velocities \cite{Hannay,CFM,LeoPiw} and like particles
in gravitational fields \cite{LeoPiw,Gordon,LeoGeo}
in general.}
moving in the potential (\ref{eq:potential}).
The frequency $\omega$ plays the role of the Hamiltonian
and the wavevector $\mathbf{k}$ corresponds to the
canonical momentum,
%%%%%%
\begin{equation}
\omega = \frac{ck}{n} \,, \quad
k = |\mathbf{k}| \,.
\label{eq:hamiltonian}
\end{equation}
%%%%%%
Indeed,
we obtain from Hamilton's equations \cite{LL1} the relations
%%%%%%
\begin{equation}
\frac{\mathrm{d}\mathbf{r}}{\mathrm{d}t} 
=\frac{\partial\omega}{\partial\mathbf{k}}
= \frac{c}{n}\,\frac{\mathbf{k}}{k} 
= \frac{c}{n^2\omega}\,\mathbf{k} 
\,,\quad
\frac{\mathrm{d}\mathbf{k}}{\mathrm{d}t}
=-\frac{\partial\omega}{\partial\mathbf{r}}
= \frac{ck}{2n^3}\,\nabla n^2
= \frac{\omega}{2n^2}\,\nabla n^2
\label{eq:hamilton}
\end{equation}
%%%%%%
that result in the equation of motion for light rays
%%%%%%
\begin{equation}
\frac{n^2}{c}\,
\frac{\mathrm{d}}{\mathrm{d}t}\,
\frac{n^2}{c}\,
\frac{\mathrm{d}\mathbf{r}}{\mathrm{d}t} 
= \frac{\nabla n^2}{2} \,.
\label{eq:motion}
\end{equation}
%%%%%%
We can express this equation as 
Newton's second law
%%%%%%
\begin{equation}
\frac{\mathrm{d}^2\mathbf{r}}{\mathrm{d}\tau^2} 
= \frac{\nabla n^2}{2} 
\label{eq:newton}
\end{equation}
%%%%%%
with the effective time increment $\mathrm{d}\tau$
measured in spatial units and defined by
%%%%%%
\begin{equation}
c\, \mathrm{d}t = n^2 \mathrm{d}\tau \,.
\label{eq:timeincrement}
\end{equation}
%%%%%%
Equation\ (\ref{eq:motion}) also reveals
the connection to Fermat's Principle \cite{BornWolf}:
light in media with refractive index $n$
takes the shortest (or longest) optical path
where the optical path length is defined,
in Cartesian coordinates, as
%%%%%%
\begin{equation}
s = \int n \sqrt{\mathrm{d}x^2+\mathrm{d}y^2+\mathrm{d}z^2}
\,.
\label{eq:optlength}
\end{equation}
%%%%%%
To see this we use the fact that 
the modulus of the Hamiltonian velocity $v$
equals $c/n$, a simple consequence of Hamilton's equations 
(\ref{eq:hamilton}), and write
%%%%%%
\begin{equation}
\frac{n}{v}\,
\frac{\mathrm{d}}{\mathrm{d}t}\,
\frac{n}{v}\,
\frac{\mathrm{d}\mathbf{r}}{\mathrm{d}t} 
= \frac{\nabla n^2}{2} \,.
\end{equation}
%%%%%%
These are the Euler-Lagrange equations \cite{LL1} 
for the effective Lagrangian $nv$.
Hence they minimize or maximize 
the integral of $nv$ that is equal to the
optical path length (\ref{eq:optlength}),
which proves Fermat's Principle.
The phase of a light ray is given by \cite{BornWolf}
%%%%%%
\begin{equation}
\phi= \int \mathbf{k}\cdot\mathrm{d}\mathbf{r} - \omega t \,.
\end{equation}
%%%%%%
Along a ray trajectory the phase $\phi$ is constant. 
Consequently,
the phase delay $\int \mathbf{k}\cdot\mathrm{d}\mathbf{r}$
corresponds to $\omega t$.
Therefore,
the Hamiltonian time $t$ measures the true time delay 
of light caused by the refractive-index profile,
whereas the Newtonian time $\tau$ serves as a convenient
parameter to characterize the ray trajectories.

\begin{figure}[h]
\begin{center}
\includegraphics[width=20.0pc]{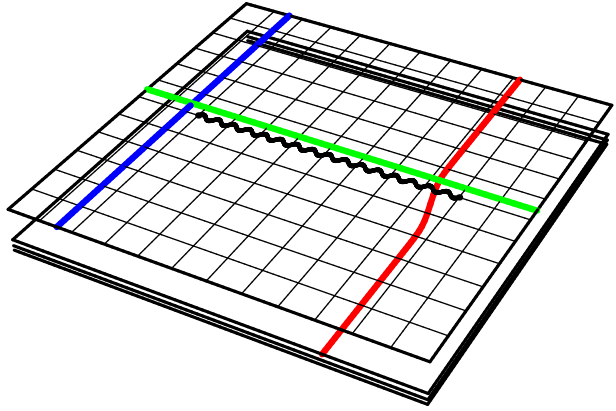}
\caption{
\small{
Optical conformal map.
A dielectric medium conformally maps physical space
described by the points $z=x+iy$ of the complex plane
onto a stack of Riemann sheets if the refractive-index profile
is $|dw/dz|$ with some analytic function $w(z)$. 
An invisibility device \cite{LeoConform}
consists of two layers in real space,
as Fig.\  \ref{fig:flow} indicates.
On the Riemann surface, 
the top sheet corresponds to the exterior
and the first lower sheet to the interior layer.
The figure illustrates the typical fates of light rays
in such media.
On the Riemann sheets rays propagate along straight lines.
The rays shown in blue and green avoid the branch cut
and hence the interior of the device.
The ray shown in red crosses the cut and passes onto the 
first lower sheet where it approaches $\infty$.  
However, this $\infty$ corresponds to a singularity
of the refractive index and not to the $\infty$ of
physical space. 
Rays like this one would be absorbed, 
unless they are guided back to the exterior sheet.}
\label{fig:sheets}}
\end{center}
\end{figure}

\begin{figure}[h]
\begin{center}
\includegraphics[width=20.0pc]{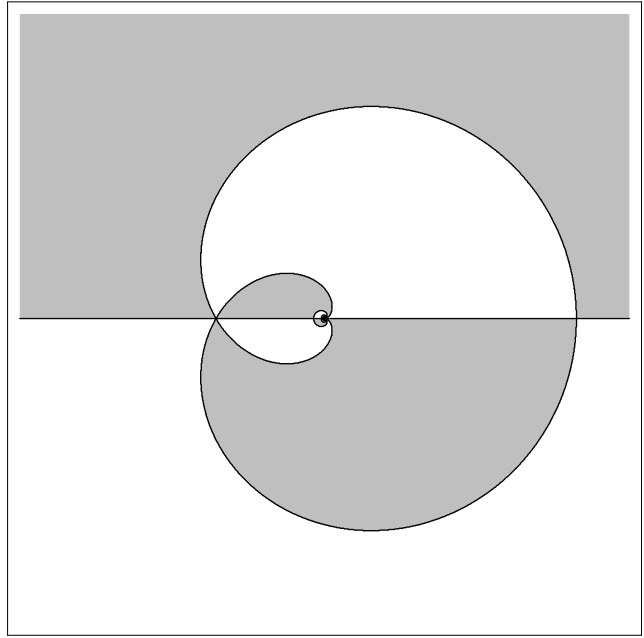}
\caption{
\small{
Riemann  sheets are tiles.
In optical conformal mapping, 
Riemann sheets represent regions of two-dimensional space,
tiles of various forms \cite{Ablowitz,Nehari}.
The figure illustrates the tiling behind the
light propagation of Fig.\ \ref{fig:flow}
(details given in the Appendix).
The upper imaginary half-plane of each sheet corresponds to a
grey tile and the lower half-plane to a white tile.
The exterior and the interior sheets of the invisibility device
occupy two of such pairs of tiles
and the hidden core takes the rest.
}
\label{fig:surface}}
\end{center}
\end{figure}

Another ingredient of our theory is optical conformal mapping
\cite{LeoConform}.
Consider an effectively two-dimensional case where
the medium is uniform in one direction and the light propagates
in a plane orthogonal to this axis. 
It is convenient to use complex numbers
$z=x+iy$ for describing
the Cartesian coordinates $x$ and $y$ in this plane.
In complex notation, the Helmholtz equation 
(\ref{eq:helmholtz}) assumes the form
%%%%%%
\begin{equation}
\left(4\frac{\partial^2}{\partial z^*\partial z} + 
n^2\frac{\omega^2}{c^2} \right) \psi = 0 \,.
\label{eq:helm}
\end{equation}
%%%%%%
Suppose that the complex $z$ coordinates are transformed to new
coordinates $w$ with an analytic function $w(z)$
that does not depend on the complex conjugate $z^*$
and hence satisfies the Cauchy-Riemann differential
equations \cite{Ablowitz}.
Such analytic functions define conformal mappings
of the complex plane onto Riemann surfaces \cite{Nehari},
see, for example, Fig.\ \ref{fig:sheets}.
%%%
%%%
Since $\partial/\partial z=
(\mathrm{d}w/\mathrm{d}z)\,\partial/\partial w$
and $\partial/\partial z^*=
(\mathrm{d}w^*/\mathrm{d}z^*)\,\partial/\partial w^*$
we obtain in $w$ space 
the Helmholtz equation (\ref{eq:helm})
with the transformed refractive-index profile $n'$
that is related to $n$ as \cite{LeoConform,Luneburg}
%%%%%%
\begin{equation}
n=n'\left|\frac{\mathrm{d}w}{\mathrm{d}z}\right| \,.
\label{eq:n}
\end{equation}
%%%%%%
On the Riemann surface, light thus propagates according to the 
refractive index profile $n'$.
The strategy \cite{LeoConform} for designing an invisibility
device is to take advantage of the sheets of the Riemann surface.
Each sheet corresponds to a distinct region in physical space,
see, for example, Fig.\ \ref{fig:surface}.
The branch cuts of the Riemann surface represent the
boundaries between the various regions.
If one wishes to hide an object, one should hide it on Riemann
sheets and prevent light from entering these sheets.
To do this,
light that has ventured across a branch cut into the interior
of the device should be guided back to the exterior.
This is done by placing a refractive-index profile on the
first interior sheet in which all ray trajectories are closed 
\cite{LeoConform}.
The device thus consists of two layers, 
an outer layer that corresponds to the exterior sheet 
on the Riemann surface in $w$ space
and an inner layer that corresponds to the first
interior sheet. 
For the outer layer we require that the refractive index
approaches unity at $\infty$, the value for empty space, 
which implies for the conformal map
%%%%%%
\begin{equation}
w(z) \sim z \quad\mbox{for}\quad z\rightarrow\infty \,.
\label{eq:asymp}
\end{equation}
%%%%%%
At the boundary between the exterior and the first interior sheet
light is refracted \cite{BornWolf}
according to Snell's law\footnote{The law of refraction was
discovered by the Arabian scientist 
Ibn Sahl more than a millennium ago \cite{Rashed}.},
unless it is totally reflected \cite{BornWolf,LeoConform}.
Since refraction is reversible,
the light rays are refracted back to the original direction
of incidence when they leave the branch cut to the exterior  sheet.
Seen on the Riemann surface, 
light rays perform loops around a branch point that
guide them back to the exterior sheet, the outside layer 
of the device.
Seen in physical space, light is guided around the object
and leaves the cloaking layers of the device as if nothing
were there. 
%%%
%%%

\newpage
 
\section{Time delay}

The time delay caused by an invisibility device depends
of course on its spatial extension.
Imagine that the refractive-index profile $n(\mathbf{r})$
is replaced by $n(\xi\mathbf{r})$ with the constant scale $\xi$.
If $\mathbf{r}(t)$ and $\mathbf{k}(t)$ are solutions of
Hamilton's ray equations (\ref{eq:hamilton}) then
$\xi\mathbf{r}$ and $\mathbf{k}$  are solutions, too, if
$t$ is replaced by $\xi t$.
This proves that the time delay is directly proportional to
the spatial extension of the refractive-index profile,
as one would expect.

Since the conformal mapping $w(z)$ 
is simply a coordinate transformation,
the time delay between two points in $z$ space 
and the delay between the corresponding points in $w$ space 
are identical.
Therefore, the part of the refractive-index profile $n$
in physical space that is due to the optical conformal mapping
does not influence the time delay at all.
Delays are only caused by the index profile in $w$
space that serves to guide light around a branch cut
on the first interior sheet.
For simplifying the notation, we denote this profile by $n$
(dropping the prime).

Suppose that the transformed refractive-index profile $n$
on the Riemann sheet is radially symmetric 
with respect to one branch point
and designed such 
that all trajectories of light rays are closed curves around that point.
The time delay $t_0$ of the invisibility device is equal to the time 
light takes to perform a loop around the branch point.
A branch point where $\nu$ sheets meet requires $\nu$ turns
($\nu$ is the winding number).
To calculate the delay,
we use polar coordinates $r$ and $\varphi$
centered at the branch point.
We obtain from the conservation law of energy \cite{LL1}
for the Newtonian dynamics (\ref{eq:newton}) of light rays
%%%%%%
\begin{equation}
\left(\frac{\mathrm{d}r}{\mathrm{d}\tau}\right)^2 + 
r^2\left(\frac{\mathrm{d}\varphi}{\mathrm{d}\tau}\right)^2
= 2\left(E-U\right) = n^2(r) \,.
\label{eq:energy}
\end{equation}
%%%%%%
As in the standard theory of motion in central potentials \cite{LL1}
we also use the conservation of the angular momentum,
%%%%%%
\begin{equation}
b = r^2 \frac{\mathrm{d}\varphi}{\mathrm{d}\tau} \,,
\label{eq:am}
\end{equation}
%%%%%%
written here in terms of the impact parameter $b$,
and obtain from the conservation laws (\ref{eq:energy}) and 
(\ref{eq:am}) the relation \cite{Luneburg}
%%%%%%
\begin{equation}
\mathrm{d}\varphi 
= \frac{b\,\mathrm{d}r}{r\sqrt{n^2r^2-b^2}} \,.
\label{eq:phi}
\end{equation}
%%%%%%
To calculate the time delay, we 
express the Hamiltonian time increment 
(\ref{eq:timeincrement}) in terms of $\mathrm{d}\varphi$,
utilizing the conservation of the angular momentum (\ref{eq:am}).
Then we use the relation (\ref{eq:phi}) to write
the time delay as an integral over the radial range of the trajectory.
The range of $r$ 
is bounded by the radial turning points $r_\pm$
where $\mathrm{d}r/\mathrm{d}\varphi$ vanishes,
which implies
%%%%%%
\begin{equation}
n^2(r_\pm)r_\pm^2 = b^2 \,.
\label{eq:turn}
\end{equation}
%%%%%%
One trajectory between $r_-$ and $r_+$ corresponds to half a turn
around the branch point.
Consequently,
%%%%%%
\begin{equation}
\frac{ct_0}{2\nu} 
= \int_{r_-}^{r_+}\frac{n^2r\,\mathrm{d}r}{\sqrt{n^2r^2-b^2}}
= \int_{r_-}^{r_+}\sqrt{n^2r^2-b^2}\,\frac{\mathrm{d}r}{r}
+ b\varphi \,.
\label{eq:t0}
\end{equation}
%%%%%%
In general, the time delay depends on the impact parameter.
However, for closed loops, $t_0$ turns out to be independent of $b$.
To see this, we differentiate the time delay (\ref{eq:t0})
with respect to the impact parameter $b$
and obtain from Eqs.\ (\ref{eq:phi}) and (\ref{eq:turn})
%%%%%%
\begin{equation}
c\,\frac{\mathrm{d}{t_0}}{\mathrm{d}b} 
= 2\nu b\, \frac{\mathrm{d}{\varphi}}{\mathrm{d}b} = 0 \,,
\end{equation}
%%%%%%
because, when all loops around the branch point are closed, 
$\varphi$ reaches $\pi$ regardless of the value of $b$.
Consequently, 
the time delay does not depend on the impact parameter
at which light has entered the branch cut to the 
first interior sheet, {\it i.e.} the interior layer of the device. 
The invisibility device causes a uniform time delay.

\section{Examples}

Reference \cite{LeoConform} mentions
two examples of refractive-index profiles on the interior sheet
that can be used to circumnavigate the branch point
such that all loops are closed,
the harmonic-oscillator profile 
%%%%%%
\begin{equation}
n_1 = \sqrt{1- r^2/r_1^2}
\label{eq:hookeforce}
\end{equation}
%%%%%%
that is related to
a Luneburg lens \cite{Luneburg,KerkerScattering} 
and the Kepler profile \cite{Luneburg,KerkerScattering}
%%%%%%
\begin{equation}
n_2 = \sqrt{r/r_2-1}
\label{eq:newtonforce}
\end{equation}
%%%%%%
that is related to an Eaton lens 
\cite{KerkerScattering,Eaton,HH}.
Here $r_1$ and $r_2$ are constants that describe the
boundaries of the refractive-index profiles
where $n$ vanishes.
Seen on the Riemann surface, 
light cannot penetrate the outside of circles of radii $r_1$
and $r_2$, respectively,
because here the refractive index would be purely imaginary.
The optical conformal mapping turns
areas on the first interior sheet inside out \cite{Ablowitz,Nehari}.
Therefore, the exterior of these circles 
corresponds to the invisible interior of the device.

The harmonic-oscillator and the Kepler potential are the
only spherically symmetric potentials $U$ where
the trajectories for all bound-state energies $E$
are closed \cite{LL1}.
However, what matters in the propagation of light rays
is the difference (\ref{eq:potential}) 
between $U$ and $E$.
Therefore it is sufficient when for a specific value of $E$
the trajectories for all angular-momenta $b$ are closed.
A known example where this is the case is Maxwell's fish eye 
\cite{BornWolf,Luneburg,KerkerScattering,Maxwell}
with the refractive-index profile
%%%%%%
\begin{equation}
n_3 = \frac{n_0}{1+(r/r_3)^2} \,.
\label{eq:maxwelleye}
\end{equation}
%%%%%%
The constant radius $r_3$ characterizes the
scale of the index profile, the radial halfpoint, and 
$n_0$ is a parameter that defines the refractive index
at the branch point.
If Maxwell's fish eye is employed to guide light
back to the exterior Riemann sheet, 
the entire interior sheet is reached by the incident light.
However,
when the Riemann surface contains more
sheets than the exterior and the first interior sheet,
all the remaining sheets are hidden.
Anything placed there is invisible. 

The Newtonian equation of motion for rays 
generated by the harmonic-oscillator profile 
(\ref{eq:hookeforce}) describes Hooke's law 
of a force proportional to the distance,
whereas the Kepler profile  (\ref{eq:newtonforce})
generates Newton's inverse-square law.
In both cases, the trajectories form ellipses
for all bound states,
a fact that Newton found exceptionally remarkable 
\cite{Principia,Chandrasekhar}.
However, Hooke's law and Newton's law can be transformed
into each other\footnote{
Ironically, despite Newton and Hooke 
reportedly having been bitter rivals,
their most celebrated force laws
are essential identical
\cite{Arnold,Needham1,Needham2}.}
by a transmutation of force 
according to the Arnol'd-Kasner theorem
\cite{Arnold,Needham1,Needham2}.

\subsection{Hooke's force}

\begin{figure}[h]
\begin{center}
\includegraphics[width=17pc]{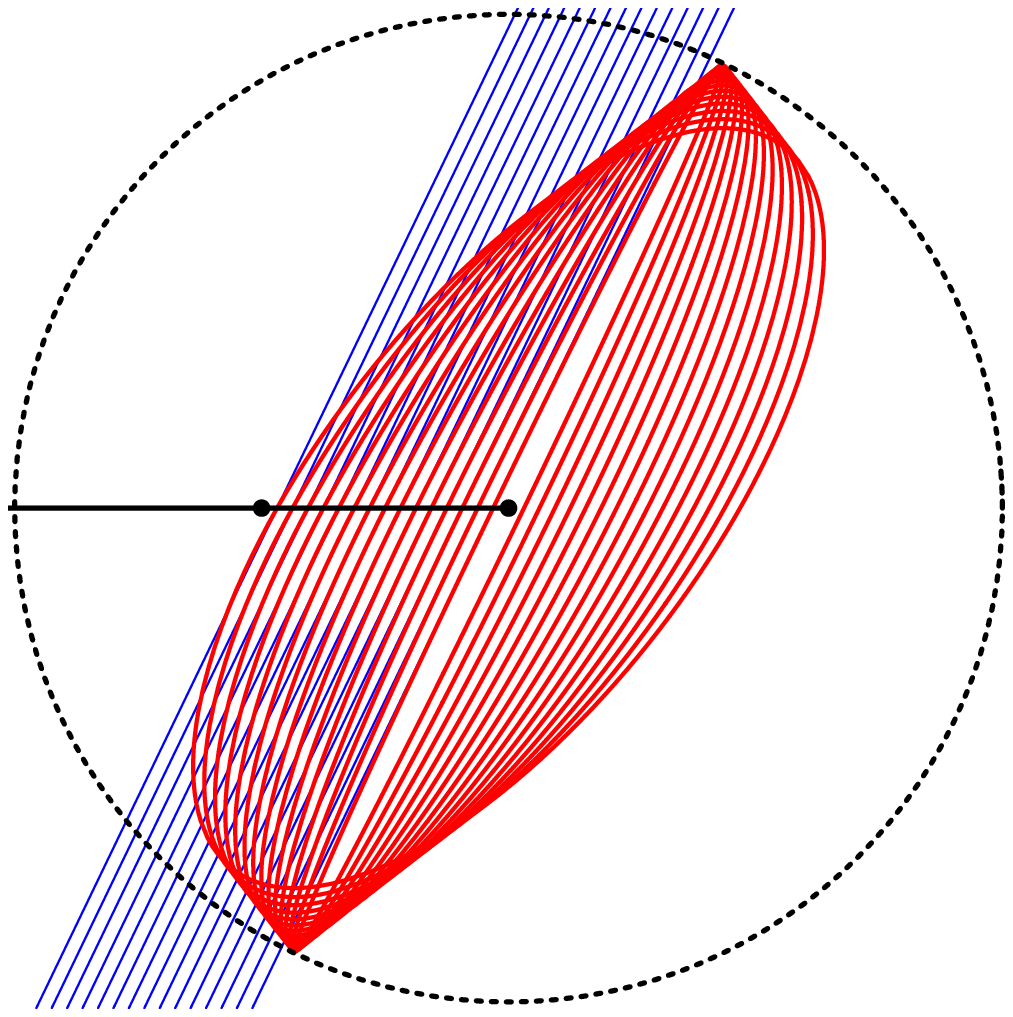}
\includegraphics[width=17pc]{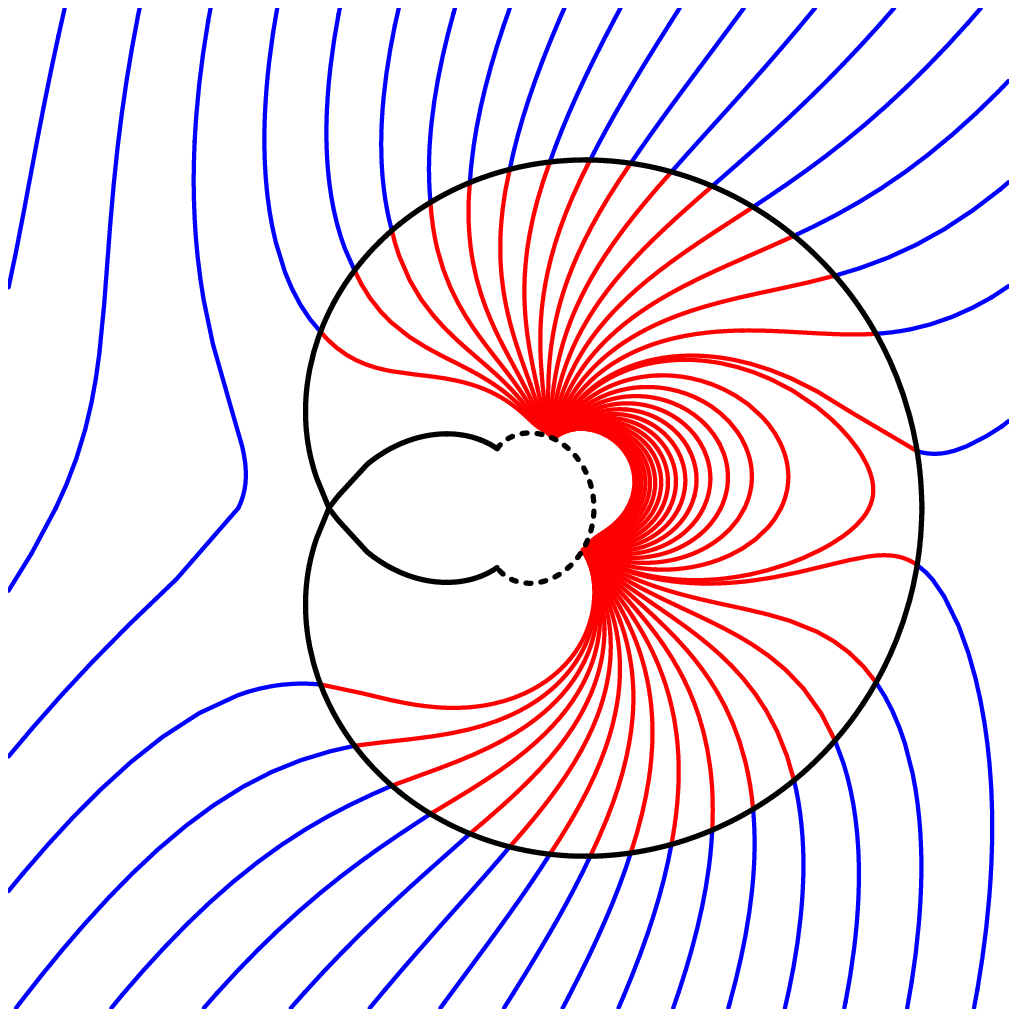}
\vspace*{5mm}
\caption{
\small{
Light guiding using Hooke's law
(similar to a Luneburg lens)
with the refractive-index profile (\ref{eq:hookeforce})
in $w$ space.
The device guides light that has entered its interior layer
back to the exterior,
represented in the left figure using two Riemann sheets that
correspond to the two layers, seen from above.
The right figure shows the corresponding ray propagation
in physical space with the optical conformal map 
(\ref{eq:wmap}) and $r_1=8r_0$.
At the branch cut in the left figure, 
the thick black line between
the two points, the branch points,
light passes from the exterior to the interior sheet.
Here light is refracted according to Snell's law,
see Eq.\  (\ref{eq:snell}) of the Appendix.
On the lower sheet, the refractive-index profile
(\ref{eq:hookeforce}) guides the rays to the exterior sheet
in elliptic orbits with one branch point in the centre.
Finally, the rays are refracted back to their original directions
and leave on the exterior sheet as if nothing had happened. 
The dotted circle in the figure indicates 
the maximal elongations of the ellipses.
This circle limits the region in the
interior of the device that light does not enter. 
The outside of the circle and the other Riemann sheets
of the map (\ref{eq:wmap}) 
correspond to the inside of the device in physical space,
as the right figure shows.
Anything inside this area is invisible.
}
\label{fig:hooke}}
\end{center}
\end{figure}

In the case of the harmonic-oscillator profile (\ref{eq:hookeforce})
the ray trajectories are very simple \cite{LL1}: 
they form a set of ellipses centered at the origin (the branch point),
see Fig.\ \ref{fig:hooke}. 
%%%
%%%
To calculate the time spent along a given ellipse
we use Cartesian coordinates $x$ and $y$
rotated such that they match the axes of this ellipse.
We describe the ellipse as
%%%%%%
\begin{equation}
x = a\cos\xi \,,\quad y = b\sin\xi \,,\quad \xi = \tau/r_1
\label{eq:ellipse}
\end{equation}
%%%%%%
with the constants $a$ and $b$ being the axis lengths.
One easily verifies that the trajectory (\ref{eq:ellipse})
solves the Newtonian equation of motion (\ref{eq:newton}).
The ray trajectory corresponds to the Newtonian energy $1/2$,
which implies
%%%%%%
\begin{equation}
2E=\left(\frac{\mathrm{d}x}{\mathrm{d}\tau}\right)^2 +
\left(\frac{\mathrm{d}y}{\mathrm{d}\tau}\right)^2 +
\frac{x^2+y^2}{r_1^2} = 
\frac{a^2+b^2}{r_1^2}=1 \,.
\end{equation}
%%%%%%
Consequently, we obtain for the time delay
%%%%%%
\begin{equation}
ct_0 
= \oint n^2 d\tau
= \nu r_1 \int_0^{2\pi}
\left(1-\frac{a^2}{r_1^2}\,\cos^2\xi - 
\frac{b^2}{r_1^2}\,\sin^2\xi \right) \mathrm{d}\xi 
= \nu\pi r_1 \,.
\label{eq:hookedelay}
\end{equation}
%%%%%%
In agreement with our general results, the time delay
caused by the harmonic-oscillator profile 
depends on the spatial extension $r_1$ and is 
otherwise uniform. 

\subsection{Newton's force}

\begin{figure}[h]
\begin{center}
\includegraphics[width=17.0pc]{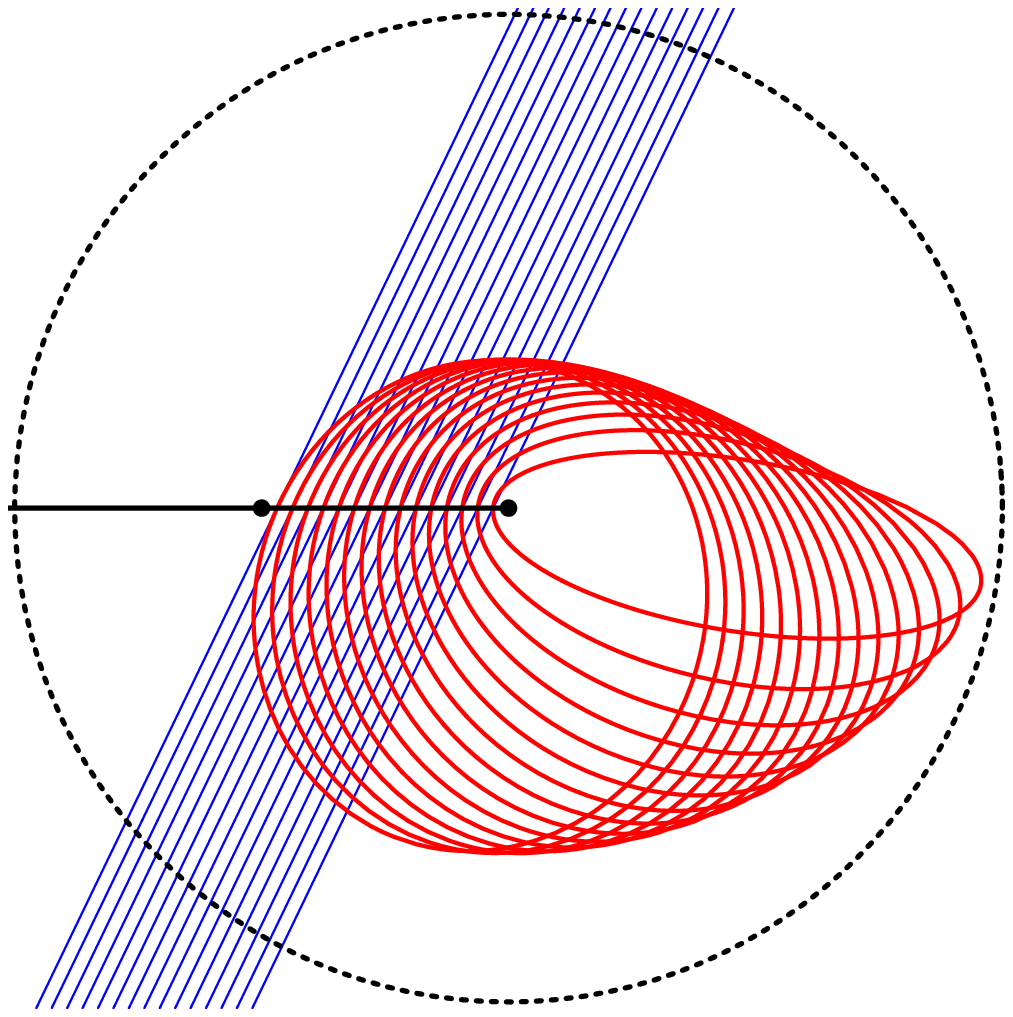}
\includegraphics[width=17.0pc]{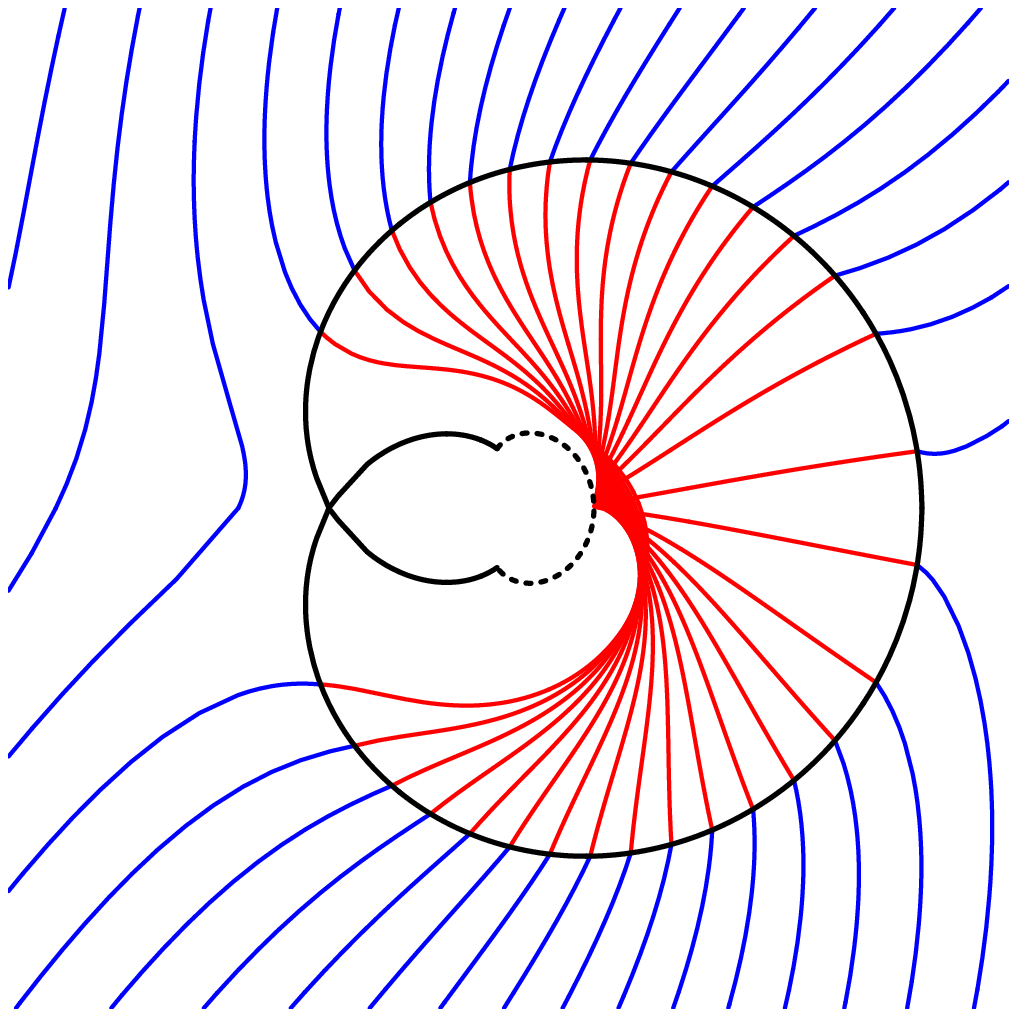}
\caption{
\small{
Light guiding using Newton's law
(related to an Eaton lens)
similar to Fig.\ \ref{fig:hooke},
with the refractive-index profile (\ref{eq:newtonforce})
in $w$ space and the optical conformal map 
(\ref{eq:wmap}) where $r_2=8r_0$.
The left figure shows how light is guided on a Riemann sheet,
whereas the right figures shows the ray propagation in physical space.
On the Riemann sheet, light rays form
elliptic orbits with one branch point in the focal point,
instead of the centre, as in Hooke's case.
}
\label{fig:newton}}
\end{center}
\end{figure}

In order to calculate the time delay caused by the Kepler profile
(\ref{eq:newtonforce}),
see Fig.\ \ref{fig:newton},
we use the transmutation of force
\cite{Arnold,Needham1,Needham2} 
to the harmonic-oscillator profile,
{\it i.e.} the transformation of Newton's law into Hooke's law
that is also based on conformal mapping \cite{Nehari}.
%%%
%%%
Consider trajectories in the complex plane, say the $z$ plane,
although in our case this plane is one of the Riemann sheets
generated by the optical conformal mapping in the first place.
Suppose that the trajectories are conformally mapped 
by the analytic function $w(z)$.
We obtain from Eqs.\ (\ref{eq:potential})  and  (\ref{eq:n})
the relations
%%%%%%
\begin{equation}
U-E = -\frac{n^2}{2} 
= -\frac{n'^2}{2}\,\left|\frac{\mathrm{d}w}{\mathrm{d}z}\right|^2
= (U'-E')\,\left|\frac{\mathrm{d}w}{\mathrm{d}z}\right|^2 \,.
\end{equation}
%%%%%%
Consequently, if the potential $U$ can be written as 
the modulus square of an analytic function,
the potential $U'$ is proportional 
to the modulus square of the inverse of this function 
expressed in terms of the new coordinates,
%%%%%%
\begin{equation}
U(z) = -E' \,\left|\frac{\mathrm{d}w(z)}{\mathrm{d}z}\right|^2
\,,\quad
U'(w) = -E \,\left|\frac{\mathrm{d}z(w)}{\mathrm{d}w}\right|^2 \,.
\end{equation}
%%%%%%
The trajectories are mapped onto each other 
by the transformation $w(z)$.
The potentials $U$ and $U'$ 
are thus related to each other, 
generating dual forces \cite{Arnold}.
Consider 
%%%%%%
\begin{equation}
w=\frac{z^2}{2r_1}  \,,\quad
E' = -\frac{1}{2} \,.
\label{eq:map}
\end{equation}
%%%%%%
The map $w(z)$ corresponds to the Hooke potential 
(\ref{eq:potential}) of the harmonic-oscillator profile
(\ref{eq:hookeforce}) with energy $E=1/2$
and, in turn, $w(z)$ generates the Kepler profile 
(\ref{eq:newtonforce}) that corresponds to
Newton's inverse square law with the parameter
%%%%%%
\begin{equation}
r_2 = \frac{r_1}{2} \,.
\end{equation}
%%%%%%
Since conformal mapping does not influence the time delay
in light propagation,
the delay generated by the Kepler profile (\ref{eq:newtonforce})
corresponds to that of the harmonic-oscillator profile  
(\ref{eq:hookeforce}), apart from one subtlety:
already a half-ellipse of the harmonic-oscillator
is mapped onto a complete Kepler ellipse, 
because of the square map (\ref{eq:map}).
Consequently, the time delay is
%%%%%%
\begin{equation}
ct_0= \frac{\nu}{2}\pi r_1 = \nu\pi r_2 \,,
\label{eq:newtondelay}
\end{equation}
%%%%%%
in complete analogy to the result (\ref{eq:hookedelay})
for the harmonic-oscillator profile.
The time delays are thus identical for identical ranges 
$r_1$ and $r_2$ of the refractive-index profiles.

\subsection{Maxwell's fish eye}

\begin{figure}[h]
\begin{center}
\includegraphics[width=18.0pc]{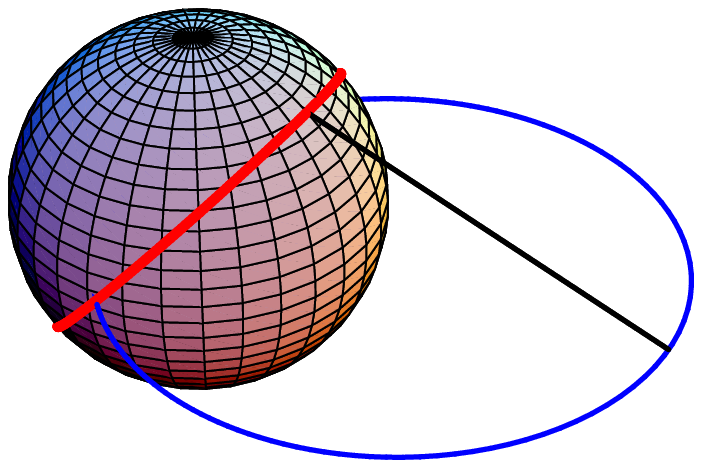}
\includegraphics[width=17.0pc]{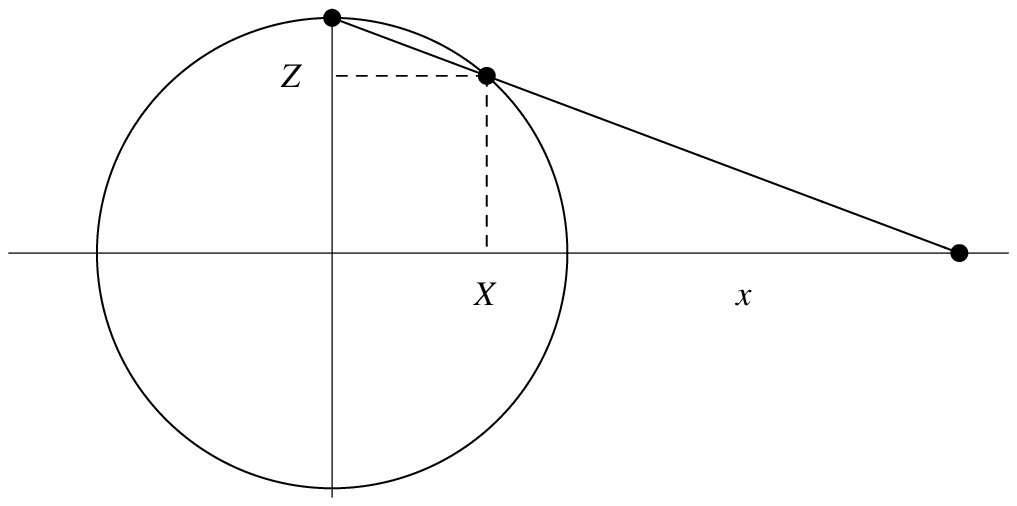}
\caption{{\small
Stereographic projection,
mapping the $(x,y)$
plane onto the  $(X,Y,Z)$ surface of a sphere.
A line drawn from the North Pole of the sphere 
to $(x,y)$ cuts the surface of the sphere at $(X,Y,Z)$.
Circles on the plane are mapped into circles 
on the sphere and vice versa \cite{Needham2}.
}
\label{fig:stereo}
}
\end{center}
\end{figure}

\begin{figure}[h]
\begin{center}
\includegraphics[width=17.5pc]{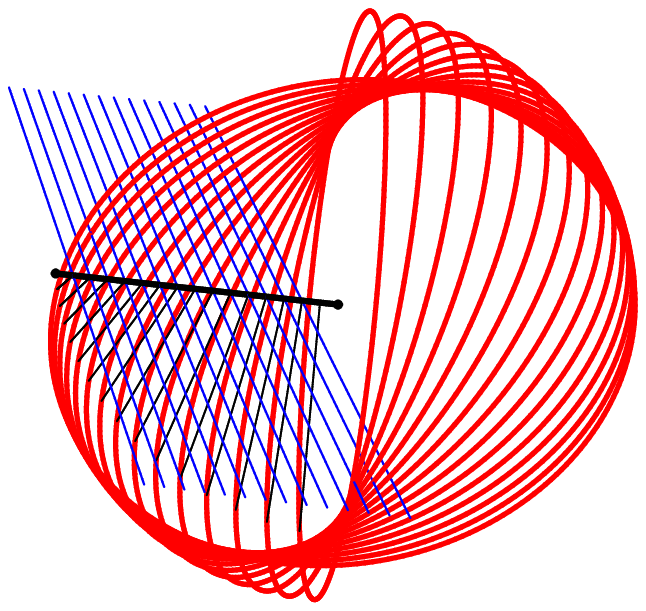}
\includegraphics[width=17.5pc]{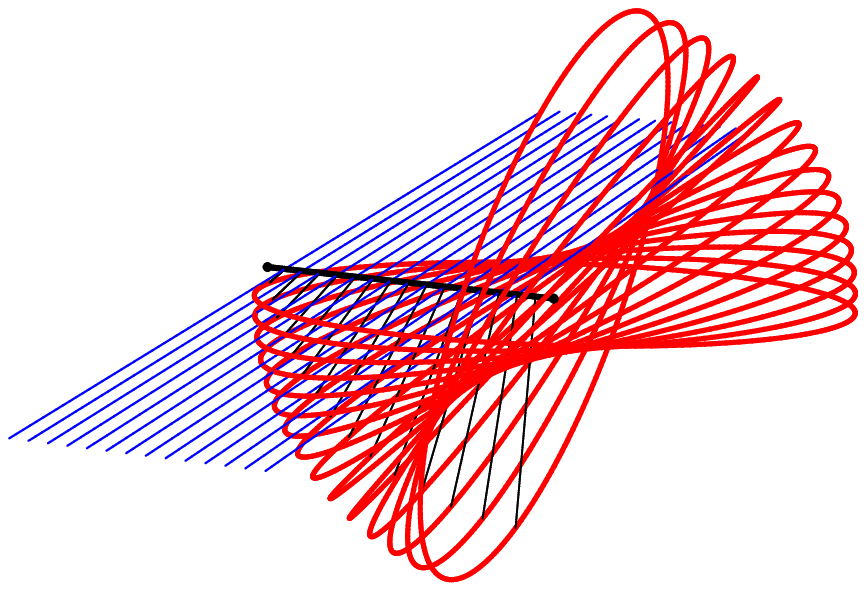}
\caption{
\small{
Light guiding using Maxwell's fish eye.
The interior layer of the invisibility device is represented
by a sphere of radius $r_3$
using the inverse stereographic map (\ref{eq:inverse}).
At the boundary, the branch cut, the light drops onto the
sphere where it propagates in great circles.
After jumping up to the exterior sheet the light rays
leave the device.
This behavior is generated by the refractive-index profile of
Maxwell's fish eye (\ref{eq:maxwelleye})
that represents 
the stereographic projection illustrated in Fig.\ \ref{fig:stereo}
as an optical conformal mapping.
The pictures illustrates light propagation for angles of incidence
of $\pm \pi/7$.
}
\label{fig:fish}}
\end{center}
\end{figure}

Maxwell's fish eye turns out to represent another classic
conformal mapping \cite{Luneburg}, 
the stereographic projection
discovered by Ptolemy and applied in 
the Mercator map projection\footnote{The Mercator map projection 
is the logarithm of the stereographic projection \cite{Needham2}.}.
Figure \ref{fig:stereo} 
%%%
%%%
shows how the points
on the surface of a sphere $(X,Y,Z)$ are mapped onto a plane,
say the $z$ plane,
according to the formulas \cite{Ablowitz,Needham2} 
%%%%%%
\begin{equation}
z=x+iy=\frac{X+iY}{1-Z/r_3} \,,\quad
X^2+Y^2+Z^2 = r_3^2 
\label{eq:stereo}
\end{equation}
%%%%%%
with the inverse
%%%%%%
\begin{equation}
X+iY = \frac{2z}{1+|r/r_3|^2} \,,\quad
Z = r_3\frac{|r/r_3|^2-1}{|r/r_3|^2+1} \,,\quad
r^2 = x^2 +y^2
\,.
\label{eq:inverse}
\end{equation}
%%%%%%
We find that the square of the
optical-length element (\ref{eq:optlength})
for Maxwell's fish eye (\ref{eq:maxwelleye}) is
%%%%%%
\begin{equation}
\mathrm{d}s^2
=n_0^2\,\frac{\mathrm{d}x^2 + \mathrm{d}y^2}{(1+|r/r_3|^2)^2}
= \frac{n_0^2}{4}\,
\left(\mathrm{d}X^2+\mathrm{d}Y^2+\mathrm{d}Z^2\right)
\,.
\end{equation}
%%%%%%
Consequently, the light rays of the fish eye are mapped into
rays on the surface of a sphere with radius $r_3$ and 
uniform refractive index $n_0/2$.
According to Fermat's Principle \cite{BornWolf}, the rays
are geodesics, lines of extremal optical path length,
see  Fig.\ \ref{fig:fish}.
On a uniform sphere, the geodesics are the great circles.
Since they are closed curves on the sphere,
the light rays are closed on the plane as well,
as we required.
Furthermore,
the stereographic projection maps circles onto circles
\cite{Needham2}
and so
the light rays in Maxwell's fish eye (\ref{eq:maxwelleye})
form circles in the plane
\cite{BornWolf,Luneburg,KerkerScattering}.
%%%
%%%

The calculation of the time delay of light circling in Maxwell's fish eye
is elementary now, because
$t_0$ does not depend on conformal transformations
and in particular on the stereographic projection (\ref{eq:stereo});
$t_0$ simply is the time delay of light during $\nu$ loops on
the surface of a sphere 
with radius $r_3$ and refractive index $n_0/2$, 
which gives
%%%%%%
\begin{equation}
ct_0 = \nu\,\frac{n_0}{2}\,2\pi r_3 = \nu\pi r_3 n_0 \,.
\end{equation}
%%%%%%
In agreement with our general results, 
the time delay is uniform and proportional to the length scale 
of the refractive-index profile.
 
\section{Conclusions}

In isotropic media,
no illusion is perfect due to the wave nature of light \cite{Nachman}.
Consequently, 
conformal invisibility devices \cite{LeoConform}
cannot be perfect; they cause reflections and time delays.
However, 
the reflectivity can be made exponentially small 
for macroscopic devices and the time delay
is uniform for all directions and impact parameters.
This is important, because images
consist of light propagating in a range of directions,
having a range of spatial Fourier components.
The time delay occurs when the light reaches
the interior layer of the device.
It will cause wavefront dislocations at the two sides,
unless the phase delay is tuned to be a multiple of $2\pi$
for some certain frequencies.
The diffraction of light will slightly blur the image
\cite{BornWolf}, but the haze caused is uniform. 

\section*{Acknowledgments}

Many people have contributed to my obsession with invisibility.
I am particularly grateful to 
Greg Gbur for our discussions in Kiev 
on the impossibility of invisibility 
and for his exquisite review article,
to Mark Dennis 
for introducing me to 
the transmutation of force and to Awatif Hindi
for her advice on elliptic modular functions.
My work has been supported by the 
Leverhulme Trust and the
Engineering and Physical Sciences Research Council.

\renewcommand{\theequation}{A\arabic{equation}}
\setcounter{equation}{0}

\section*{Appendix}

\begin{figure}[h]
\begin{center}
\includegraphics[width=25.0pc]{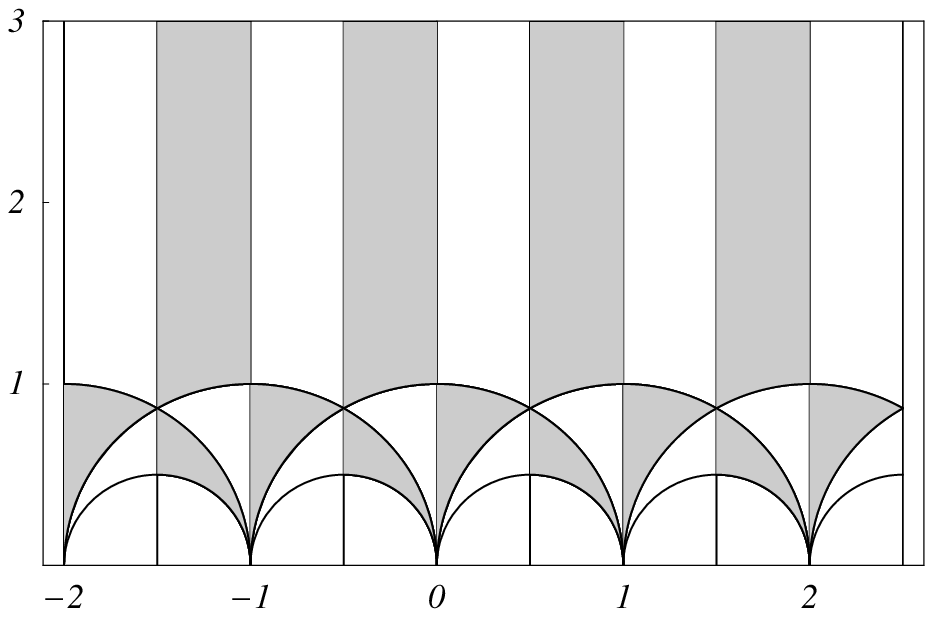}
\caption{
\small{
The Klein invariant $J(z)$ tiles the upper half $z$ plane
with an infinite sequence of circular arches.
The arches continue near the real axis
in an infinitely intricate structure (not shown here). 
Identifying all horizontal strips of length $1$
and deforming the outer arch to a circle leads to
Fig.\ \ref{fig:surface} that illustrates the tiling of
the optical conformal map (\ref{eq:wmap}).
}
\label{fig:klein}}
\end{center}
\end{figure}

Most pictures of this paper are based on a conformal map using
elliptic modular functions \cite{Erdelyi}.
These functions are connected to many branches of mathematics,
including the proof of Fermat's Last Theorem \cite{Wiles}.
We adopt the notation of the Bateman Manuscript Project
on Higher Transcendental Functions \cite{Erdelyi}
(not Neharis notation\footnote{In Nehari's book
\cite{Nehari} $\lambda(z)$ is denoted as $J(z)$.}).
We use the modular function $J$ 
known as the Klein invariant,
illustrated in Fig.\ \ref{fig:klein},
and expressed here in terms of the modular function $\lambda$ as
%%%%%%
\begin{equation}
J(z) = \frac{4}{27}\,
\frac{(1-\lambda+\lambda^2)^3}{\lambda^2(1-\lambda)^2}
\,,\quad
\lambda = 16q \left(
\frac{\displaystyle \sum_{m=0}^\infty q^{m(m+1)}}
{\displaystyle 1 + 2 \sum_{m=1}^\infty q^{m^2}}
\right)^4
\label{eq:j}
\end{equation}
%%%%%%
with
%%%%%%
\begin{equation}
q = e^{i\pi z}\,,\quad |q|<1 \,,
\end{equation}
%%%%%%
see Eqs.\ (36-38) of Ref. \cite{Erdelyi}.
%%%
%%%
Note that the expression for $\lambda$ is rapidly converging
and therefore well suited for numerical computations.
Consider the map
%%%%%%
\begin{equation}
w = 4r_0 J\left(-\frac{\ln(432z/r_0)}{2\pi i}\right)
- \frac{31r_0}{18}
= \frac{16r_0}{27}\,
\frac{(1-\lambda+\lambda^2)^3}{\lambda^2(1-\lambda)^2}
- \frac{31r_0}{18}
\label{eq:wmap}
\end{equation}
%%%%%%
with
%%%%%%
\begin{equation}
q = \frac{1}{\sqrt{432z/r_0}} \,.
\label{eq:q2map}
\end{equation}
%%%%%%
The constant $r_0$ 
characterizes the spatial scale of the optical conformal mapping.
Far away from the device light should propagate through
empty space, which implies that $w\sim z$ for $z\rightarrow\infty$.
The map (\ref{eq:wmap}) is chosen such that 
this is the case.
Indeed, we obtain from the representation (\ref{eq:j})
the first terms of the Laurent expansion
%%%%%%
\begin{equation}
J \sim \frac{1}{1728 q^2} + \frac{31}{72} 
+ \frac{1823}{16}\,q^2 
\quad\mbox{for}\quad q\rightarrow 0 \,,
\end{equation}
%%%%%%
which implies, according to Eqs.\ (\ref{eq:wmap})
and (\ref{eq:q2map}), 
%%%%%%
\begin{equation}
w \sim z + \frac{1823}{1728}\,\frac{r_0^2}{z}
\quad\mbox{for}\quad z\rightarrow\infty \,.
\label{eq:asym}
\end{equation}
%%%%%%
In the exterior of the device, the map (\ref{eq:wmap}) approaches 
the simple example considered in Ref.\ \cite{LeoConform},
whereas in the interior the map (\ref{eq:wmap})
represents the infinitely more complicated Riemann surface
illustrated in Figs.\ \ref{fig:surface} and \ref{fig:klein}. 
The Riemann surface contains three branch points 
\cite{Nehari,Erdelyi},
$w_1=(41/18)r_0$, $w_2=-(31/18)r_0$ and $w_\infty=\infty$
with winding numbers $1$, $2$ and $\infty$,
apart from the exterior sheet with only $w_1$ and $w_2$
\cite{Nehari,Erdelyi}.

In order to calculate the ray trajectories, we consider
the ray dynamics in $w$ space and then transform $w$
to the physical trajectories in $z$ space.
We describe both the ray trajectories $w$
and the wavevectors $k$ by complex numbers.
In the exterior sheet light propagates along straight lines.
Given a point $w$ on the exterior sheet,
we numerically solve Eq.\ (\ref{eq:wmap}) for $z$
using the inversion $z(w)$ 
of the asymptotic map (\ref{eq:asym})
as the starting value,
%%%%%%
\begin{equation}
z \sim \frac{1}{72}\left(36w+\sqrt{3(432w^2-1823)}\right)
\end{equation}
%%%%%%
for $\mathrm{Im}w \ge 0$ and we use $z^*(w^*)$ otherwise.
At the branch cut between the exterior and the first interior sheet light
is refracted according to Snell's law \cite{BornWolf,Rashed}.
Since the modulus of $k$ equals $n(\omega/c)$,
we obtain for a light ray incident at the angle $\varphi$
Snell's law in complex notation as
%%%%%%
\begin{equation}
k = \frac{\omega}{c}
\left(\sin\varphi - i \sqrt{n^2-\sin^2\varphi}\right) \,.
\label{eq:snell}
\end{equation}
%%%%%%
On the first interior sheet, we solve Hamilton's equations
for a radially symmetric index profile around the
branch point $w_1$,
%%%%%%
\begin{equation}
\frac{\mathrm{d}w}{\mathrm{d}l} = \frac{k}{n(|w-w_1|)\,|k|}
\,,\quad
\frac{\mathrm{d}k}{\mathrm{d}l} =
\frac{n_r(|w-w_1|)\,|k|}{n^2(|w-w_1|)}\,\frac{w}{|w|}
\,,\quad
n_r(r)=\frac{\mathrm{d}n(r)}{\mathrm{d}r}
\end{equation}
%%%%%%
for the propagation distance $l=ct$
and Maxwell's fish eye (\ref{eq:maxwelleye})
with $n=n_3$, $r_3=4r_0$ and $n_0=2$.
The parameters are designed such that the refractive index
on the Riemann structure reaches 1 at the other branch point 
$w_2$ and exceeds $1$ along the branch cut.
In this way, total reflection is excluded for the lowest
possible value of $n_0$.
To calculate $z(w)$ on the first interior sheet
we utilize the modular symmetry of the Klein invariant
\cite{Erdelyi},
%%%%%%
\begin{equation}
J(z') = J(z) \quad\mbox{for}\quad z'=-z^{-1} \,,
\end{equation}
%%%%%%
which leads to 
%%%%%%
\begin{equation}
z' = \frac{r_0}{432}\,
\exp\left(\frac{4\pi^2}{\ln(432 z/r_0)}\right)
\end{equation}
%%%%%%
for the position $z'$ that shares the same numerical value
of $w$ as $z$, but corresponds to the first interior sheet. 
We follow the same procedure as for the exterior layer
of the device to calculate $z$ and then transform it into $z'$
to continue the trajectory in the interior layer.

%%%

\end{document}